\documentclass[prd,twocolumn,nofootinbib,showpacs,superscriptaddress]{revtex4}

\usepackage{amsfonts}
\usepackage{amsmath}
\usepackage{bm}
\usepackage{graphicx}
\usepackage[latin1]{inputenc}
\usepackage{latexsym}
\usepackage{rotating}

\def\newacronym#1#2#3{\gdef#1{#3 (#2)\gdef#1{#2}}}

\newacronym{\NSF}{NSF}{National Science Foundation}
\newacronym{\NASA}{NASA}{National Aeronautics and Space Administration}
\newacronym{\lisa}{LISA}{the Laser Interferometer Space Antenna}
\newacronym{\ligo}{LIGO}{Laser Interferometer Gravitational-wave Observatory} 
\newacronym{\Caltech}{Caltech}{California Institute of Technology}
\newacronym{\MIT}{MIT}{Massachusetts Institute of Technology}
\newacronym{\sph}{SPH}{smooth particle hydrodynamics}
\newacronym{\tsi}{TSI}{the Terascale Supernova Initiative}
\newacronym{\wmap}{WMAP}{the Wilkinson Microwave Anisotropy Probe}
\newacronym{\decigo}{DECIGO}{the Deci-Hertz Interferometric
  Gravitational-wave Observatory} 
\newacronym{\cmbr}{CMBR}{cosmic microwave background}
\newacronym{\ibbh}{IBBH}{intermediate binary black hole}
\newacronym{\bdj}{BDJ}{Brans-Dicke-Jordan}
\newacronym{\bbo}{BBO}{Big Bang Observer}
\newacronym{\decigo}{DECIGO}{Deci-Hertz Gravitational-Wave Observatory}

\def\MPR#1{{\it Moving Puncture Recipe}#1 (MPR#1)\gdef\MPR{MPR}}
\def\ahz#1{apparent horizon#1 (AH#1)\gdef\ahz{AH}}
\def\CLA#1{close-limit approximation#1 (CLA#1)\gdef\CLA{CLA}}
\def\pnw#1{post-Newtonian#1 (PN#1)\gdef\pnw{PN}}
\def\qnm#1{quasi-normal mode#1 (QNM#1)\gdef\qnm{QNM}}
\def\isco#1{innermost stable circular orbit#1 (ISCO#1)\gdef\isco{ISCO}}
\def\eos#1{equation of state#1 (EOS#1)\gdef\eos{EOS}}
\def\tov#1{Tolman-Oppenheimer-Volkoff#1 (TOV#1)\gdef\tov{TOV}}
\def\ns#1{Neutron Star#1 (NS#1)\gdef\ns{NS}}
\def\lqc#1{Loop Quantum Cosmology#1 (LQC#1)\gdef\lqc{LQC}}
\def\wdw#1{Wheeler-DeWitt#1 (WDW#1)\gdef\wdw{WDW}}
\def\tnw#1{Teukolsky-Nakamura wave#1 (TNW#1)\gdef\tnw{TNW}}
\def\tn#1{Teukolsky-Nakamura#1 (TN#1)\gdef\tn{TN}}
\def\bbh#1{binary black hole#1 (BBH#1)\gdef\bbh{BBH}}
\def\nr#1{numerical relativity#1 (NR#1)\gdef\nr{NR}}
\def\bhns#1{Black Hole -- Neutron Star#1 (BHNS#1)\gdef\bhns{BHNS}}
\def\nsns#1{Neutron Star -- Neutron Star#1 (NSNS#1)\gdef\nsns{NSNS}}
\def\emri#1{Extreme Mass-Ratio Inspiral#1 (EMRI#1)\gdef\emri{EMRI}}
\def\emrb#1{Extreme Mass-Ratio Binaries#1 (EMRB#1)\gdef\emrb{EMRB}} 
\def\grb#1{Gamma-Ray Burst#1 (GRB#1)\gdef\grb{GRB}}
\def\imbh#1{intermediate mass black hole#1 (IMBH#1)\gdef\imbh{IMBH}}
\def\smbh#1{supermassive black hole#1 (SMBH#1)\gdef\smbh{SMBH}}
\def\bh#1{Black Hole#1 (BH#1)\gdef\bh{BH}}
\def\ulx#1{ultra-luminous x-ray source#1 (ULX#1)\gdef\ulx{ULX}}
\def\lmxbs{low-mass x-ray Binaries (LMXBs)\gdef\lmxbs{LMXBs}\gdef\lmxb{LMXB}} 
\def\lmxb{low-mass x-ray Binary (LMXB)\gdef\lmxbs{LMXBs}\gdef\lmxb{LMXB}}

\begin{document}

\title[Robustness of Binary Black Hole Mergers in the Presence of Spurious Radiation]{Robustness of Binary Black Hole Mergers in the Presence of Spurious Radiation}
\author{Tanja Bode}
\author{Deirdre Shoemaker}\altaffiliation[Also at ]{Department of Physics and Institute for Gravitation and the Cosmos, The Pennsylvania State University, University Park, PA 16802, USA}
\author{Frank Herrmann}
\author{Ian Hinder}
\affiliation{Center for Gravitational Wave Physics, \\
The Pennsylvania State University, University Park, PA 16802, USA}
\date{\today}

\begin{abstract}
We present an investigation
into how sensitive the last orbits and merger of binary black hole systems are to the presence of
spurious radiation in the initial data.  Our numerical experiments consist of a binary black hole
system starting the last couple of orbits before merger with additional spurious radiation 
centered at the origin and fixed initial angular momentum. 
As the energy in the added spurious radiation increases, the binary is invariably hardened for the cases we tested, i.e. 
the merger of the two black holes is hastened.
The change in merger time becomes significant when the additional energy provided by the spurious
radiation increases the Arnowitt-Deser-Misner (ADM) mass of the spacetime by about 1\%.  While the final
masses of the black holes increase due to partial absorption of the radiation, the final spins
remain constant to within our numerical accuracy.
We conjecture that the spurious radiation is primarily increasing the eccentricity of the orbit and secondarily
increasing the mass of the black holes while propagating out to infinity.
\end{abstract}

\pacs{04.25.dg,04.30.Db,04.70.Bw}

\maketitle

\section{Introduction}\label{intro}
The coalescence of two black holes, long thought of as the ``holy grail'' of \nr{},
is well on its way to being a solved problem. 
Many groups in \nr{} have now demonstrated the ability to
follow two black holes through several orbits \cite{Pfeiffer07} and their final orbits and merger to a
single black hole~\cite{Pretorius:2005gq, Baker05a, Campanelli05a,herrmannunequal,gonzalez-2007-98,koppitz-2007,
2007PhRvD..76f1502T}.
From the first published waveform of equal-mass, non-spinning \bbh{} coalescence, 
the simplicity of the waveform's dependence
on time has been noted.  Comparisons amongst the groups in \nr{}
have demonstrated a remarkable agreement to the solution of the \bbh{} problem.
A common aspect in all numerical relativity \bbh{} evolutions is the presence of spurious radiation
in the initial data.  
In this paper, we present a study on how the standard equal-mass, quasi-circular \bbh{} system responds
to the presence of spurious radiation that has been added in a controlled manner and map that response as a function of the radiation's 
initial conditions.  Our intent is to determine how much junk radiation the system can handle
and how the waveforms and the physical properties of the final black hole deviate from the standard
\bbh{} result.

Several papers have compared \bbh{} waveforms.   One of the first papers to internally compare
waveforms also demonstrated the first evidence of ``universality''\cite{Baker06} 
in an equal-mass, non-spinning initial configuration.  The paper demonstrated that differences
in initial data characterized by a change in the initial orbital separation
manifested as a time shift in the amplitude and phase of the gravitational 
waveforms.  Once time-shifted, the waveforms were within 1\% agreement over the 
merger and ringdown in $|r_0 \Psi_4|$. 
We investigate the effect that the additional spurious radiation we add to the binary will have on this universality.

The first comparison of \nr{} waveforms between several groups~\cite{Baker07} includes the most popular methods used 
in the community to evolve \bbh{s} covering excision with a hyperbolic formulation \cite{Pretorius:2005gq}
and moving punctures with the Baumgarte-Shapiro-Shibata-Nakamura (BSSN) formulation of the Einstein equation \cite{ Baker05a, Campanelli05a}.
The waveforms were in remarkable agreement once time-shifted,
the largest differences, occurring at the beginning of the wave, being due to the spurious radiation
in the initial data.  A second, independent comparison of waveforms from different methods was conducted by Sperhake
\cite{SperhakeKS} in which he compared a Kerr-Schild/excision evolution to a puncture evolution within 
the same code.  An interesting issue to investigate is to what extent the spurious radiation in the initial
data could cause differences in the merging time and thus affect waveform comparisons
based on time-shifts to align the amplitude of the waveform.

Most groups remove the initial burst from the waveform during post-processing of the
data~\cite{Camp0601,Choi07}.
From the evolutions published, it appears that the spurious radiation that is
present in the initial data is flushed out of the system within a crossing
time, leaving the binary dynamics mostly untouched.  There
is still some concern about the impact that choices made in setting up the initial
data for the evolutions, choices like conformal flatness, 
have on the waveforms.  Studies have looked at different ways of choosing
the freely specifiable part of initial data~\cite{hannam06} which reduce the amplitude of the spurious 
radiation, but these have not been extensively implemented in evolutions.

In this paper, we test the robustness of the binary to effects of spurious radiation.  
We create  a \bbh{} system containing additional radiation with tunable initial energy 
initialized at the binary's center of mass. 
We then evolve a series of equal-mass, non-spinning, quasi-circular \bbh{} plus radiation spacetimes 
using the PSU numerical code that implements the \MPR{}~\cite{Baker05a,Campanelli05a}, 
see \cite{vaishnav-2007} for details about the PSU code.   
The initial data used to construct the \bbh{} plus radiation spacetime
is presented in Section~\ref{Initial Data}, the results in Section~\ref{Effects of TNW}
and the conclusions in Section~\ref{Conclusions}.  Our main result is
that the presence of spurious radiation causes a hastening of the merger, thus plausibly accounting for
the differences in merger times seen in the \nr{} waveform comparisons.  Before describing 
our numerical experiment, we present a 
back-of-the envelope calculation to build our intuition about this problem.

\section{A Newtonian Perspective}
For illustrative purposes, we investigate the effect a central pulse of energy might
have on a binary by studying a two-body orbit 
in Newtonian gravity with a stationary mass placed at the orbit's center of motion while
the bodies are at their apocenter.  
The addition of the third mass at the center of the Newtonian binary affects the
orbit by deepening the potential in which the binary sits.  We solve the problem using
the standard central force solution to the two-body problem with the new potential.
We assume $\dot{r}=0$ initially since this is set in the parameters to the initial data
solver.  Letting $m$ be the masses of the 
black holes and $m_w$ be the equivalent mass of the third body, we write the ratio of the
final eccentricity, $e^{\prime}$, to the original eccentricity as 
\begin{equation}
(\frac{e^{\prime}}{e})^2 = \frac{1}{(1+2f)^2} [1 + \frac{(4-2 j^2/\mu d)f + 4 f^2}{e^2} ]\,
\end{equation}
where $f=m_w / m$ is the fractional mass, $j=l/\mu$ is the angular 
momentum per unit reduced mass, and $d$ is the initial separation of the binary.  

This simple calculation indicates that the eccentricity increases for sufficiently small 
eccentricities.  For the binary parameters studied in this paper, the eccentricity 
invariably increases for $e\le0.88$.
Although the black holes in our \bbh{} evolutions are not far enough apart to 
allow a valid determination of eccentricity, the trajectories are quasi-circular enough for 
the eccentricity to be low.  This illustrates that we can expect the addition of radiation into 
the studied system to cause the binary's orbit to become more elliptic.

\section{Injecting Radiation into a BBH Evolution}\label{Initial Data}

We inject radiation into the standard, equal-mass, non-spinning, quasi-circular
\bbh{} evolution during the setup of the
initial data.  The initial data for the evolution is constructed via the puncture
method \cite{Brandt97} using
the single-domain spectral method code developed by Ansorg et
al.~\cite{Ansorg04} that uses a conformally flat prescription to solve
the constraints.  We have two building blocks for the data: 1) the
quasi-circular \bbh{} and 2) the tunable radiation.  The \bbh{} data
is set-up using the input conditions for the Baker et
al.~\cite{Baker06} R1 run of two equal-mass irrotational black
holes in quasi-circular orbits such that the metric is conformally
flat.  The details of the R1 initial data are given in the first row 
of Table~\ref{runtaball} and a convergence study was done 
in~\cite{vaishnav-2007}.

\subsection{The Teukolsky-Nakamura Wave}
The tunable radiation is given by an even parity, quadrupolar
gravitational wave: the linearized solution to a perturbation on
Minkowski spacetime expanded over the modes of the Matthews tensor
spherical harmonics \cite{RevModPhys.52.299}.  This radiation was first derived by 
Teukolsky~\cite{Teuk82} and is typically known as a Teukolsky wave.  This
tensor is simply the general traceless-transverse solution to the linearized Einstein 
equation. Nakamura~\cite{Nakamura87}  later
wrote out the solution for general $\ell$ and $m$.  Instead
of using that tensor as a metric perturbation,
he used it as an extrinsic curvature perturbation.
We implement the Nakamura version
of the Teukolsky wave, herein called \tnw{s}, 
in order to satisfy the condition of a conformally flat metric imposed by the 
puncture method.

The \tn{} extrinsic curvature tensor is given by
\begin{widetext}
\begin{equation} \label{eq:TNAij}
\tilde{A}_{ij}^\mathrm{TN}=\sum_{l,m}\left(\begin{array}{ccc}
    a_{lm}Y_{lm} & b_{lm}Y_{lm,\theta} & b_{lm}Y_{lm,\varphi} \\
     & g_{lm}Y_{lm}+f_{lm}W_{lm} & f_{lm}X_{lm} \\
     &  & (g_{lm}Y_{lm}-f_{lm}W_{lm}) \sin^2\theta 
    \end{array} \right) 
\end{equation}
where the coefficients $a_{lm}$, $b_{lm}$, $f_{lm}$, and $g_{lm}$ are functions only of $r$ 
and $t$ as follows
\begin{subequations} \label{eq:TNcoefs}
\begin{eqnarray}
 a_{lm}&=&r^{l-2}\left(\frac{1}{r}\frac{\partial}{\partial r}\right)^l \frac{F(t-r)+F(t+r)}{r}, \\
 b_{lm}&=&\frac{1}{l(l+1)r}\frac{\partial}{\partial r}(r^3 a_{lm}), \\
 g_{lm}&=&-\frac{r^2}{2}a_{lm}, \\
 f_{lm}&=&\frac{1}{(l-2)(l+1)}\left[g_{lm}+\frac{\partial}{\partial r} \left(\frac{r}{l(l+1)}
	 \frac{\partial}{\partial r}(r^3 a_{lm})\right)\right]
\end{eqnarray}
\end{subequations}
and the angular functions $X_{lm}$ and $W_{lm}$ are 
\begin{subequations} \label{eq:TNangcofs}
\begin{eqnarray}
  X_{lm}&=&2\frac{\partial}{\partial \varphi}\left(\frac{\partial}{\partial \theta}-\cot\theta\right)Y_{lm}, \\
  W_{lm}&=&\left(\frac{\partial^2}{\partial\theta^2}-\cot\theta\frac{\partial}{\partial\theta}
	  -\frac{1}{\sin^2\theta}\frac{\partial^2}{\partial\varphi^2}\right)Y_{lm}\,.
\end{eqnarray}
\end{subequations}
\end{widetext}

Note that the \tn{} solution lets us choose the radial dependence
in the form
of ingoing and outgoing functions which we have chosen to be
the same symmetric functional form, $F(u)$.
$F(u)$ is given by an Eppley packet \cite{Eppley}
\begin{equation} 
F(u) = A u e^{-u^2 / \sigma^2} \,,
\end{equation} 
where $u=t\pm r\,.$
The Eppley packet is localized and smooth with the factor of $u$ present 
so the wave is more regular at the origin.  Later in the paper we 
investigate a cosine modulation of this function.
We can choose the location, mode content, strength, and radial
dependence of the injected radiation. Since the 
extrinsic curvature is real and the spherical harmonics are
complex, we take only the real part of the $X_{lm}$  tensor
resulting in a superposition of $m$ and
$-m$ modes in the \tnw{}. 

\subsection{\bbh{}+\tnw{}}

Our initial data, the spatial metric and extrinsic curvature,
are given by 
\begin{subequations} \label{eq:Decomp}
\begin{eqnarray}
g_{ij} &=& \psi^{4} \eta_{ij}, \\
K_{ij} &=& \psi^{-2} (\tilde A^{BY}_{ij}+\tilde A^{TN}_{ij})\,,
\end{eqnarray}
\end{subequations}
where $\eta_{ij}$ is the flat spatial metric and $\psi$ is the solution to the Hamiltonian 
constraint under York's conformal approach \cite{York79}. The extrinsic curvature $\tilde A^{BY}_{ij}$ is the
Bowen and York 
solution \cite{Bowen:1980yu} to the momentum constraint and $\tilde A^{TN}_{ij}$ is the \tn{} tensor.
Notice that because the momentum constraint is linear in the extrinsic curvature, the 
superposition of the extrinsic curvature is also a solution of the momentum constraint.
As a test we evolved the above initial data with a vanishing Bowen-York tensor, i.e. Minkowski background.
For the $\ell=2$, $m=\pm2$ case, we found that the Arnowitt-Deser-Misner (ADM) angular 
momentum calculated on the initial spacetime is zero to within machine error.
The angular momentum of the \bbh{}+\tnw{} is therefore independent of the \tnw{} to our numerical accuracy.

\subsection{Configurations}
\begin{figure}[ht]
\begin{center}
\includegraphics{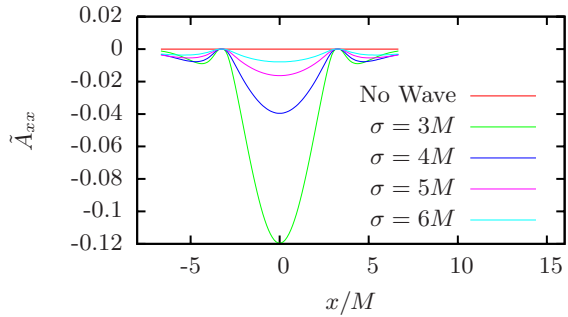}
\caption{\label{fig:idA11}Comparison of $\tilde{A}_{xx}$ initially between
various Eppley packet widths, $\sigma$, for a wave amplitude of $1 M^3$.  This
is from the binary black hole initial data where $\tilde{A}_{xx}$ vanishes 
at the punctures.}
\end{center}
\end{figure}

The simplest geometry to add additional spurious radiation to our \bbh{}
initial data is to center a wave pulse at the origin.  
We typically choose an $\ell=2$, $m=2$ mode, the dominant 
mode for gravitational radiation from a \bbh{} system,
and vary the amplitude, $A$, and width, $\sigma$, of the Eppley packet.
The values of $\sigma/M$ are chosen from the set \{0,3,4,5,6\} and those of
the amplitude $A/M^3$ from the set \{0,0.1, 0.5, 1.0, 1.5\} where the dry R1 \bbh{} 
spacetime is recovered when $A/M^3=0\,.$  
Fig.~\ref{fig:idA11} shows the
shape of the wave in one of the components of the traceless-transverse 
extrinsic curvature, $A_{ij}=\psi^{-2} \tilde{A}_{ij}$, along the
coordinate axis intersecting the two black holes, modulated by the
inverse square of the conformal factor, for $A=1 M^3$ and
$\sigma=3M,4M,5M,6M$. 

When adding the \tnw{} to the spacetime we wanted to keep the initial
black holes unaltered.  We chose to keep the \ahz{} masses constant independent of 
the additional wave content.  
In practice the \ahz{} masses varied by as much as 0.04\% 
from the dry R1 run due to insufficient parameter accuracy. The momenta remain
constant as parameters to the initial data solver, and the ADM angular momentum 
differs by at most
0.001\% from the dry R1 run.   The second column of Table~\ref{runtaball}
lists the ADM energy of the spacetime for each wave choice.  We
note that the waves increase the ADM energy from a negligible $10^{-4}\%$
to a significant 8.9\%, which scales empirically as $E_\mathrm{ADM}\simeq A^2/\sigma^5$.
The proper separation between the black holes
changes from the dry case of $L=9.94$ to a 
maximum of $L=10.23$.  In the most extreme case, the
wave having $A=1.5 M^3$ and $\sigma=3M$,  we have added almost 9\%
additional energy into the \bbh{} system and effectively moved the black holes
apart by $0.29M$. The impact of these differences in initial data on 
the binary evolution are discussed further in the next section. 

\begin{figure}[h]
\begin{center}
\includegraphics[width=80mm]{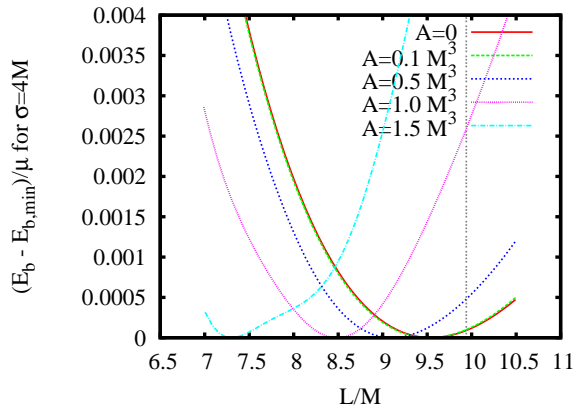}
\caption{\label{fig:Ebind}We plot the effect of the wave on the binding
energy per unit reduced mass in the initial data.  The potentials were
calculated by solving the initial data using Ansorg's code for various
separations while keeping the individual \ahz{} masses and total ADM angular momentum
fixed. }
\end{center}
\end{figure}

Pumping energy into the system while holding the coordinate separation
and angular momentum constant necessarily means that we are changing
the binding energy of the system.  To study this change
we map the effective potentials for each \bbh{}+\tnw{} case.
We do this by repeatedly solving the initial data
with incremented separations while holding the individual \ahz{} masses and
total ADM angular momentum fixed.
We calculated the quantity 
$E_b=E_\mathrm{ADM}-m_{\mathrm{AH},1}-m_{\mathrm{AH},2}$ for each spacetime.  
The wave itself adds to the ADM energy and must be subtracted in this calculation; but, 
as we're only interested in the relative shapes, we can look at the relative binding 
energy, $E_b-E_{b,\mathrm{min}}$.
For the waves with a $\sigma$ of $4M$
the binding energy per unit reduced mass is plotted in
Fig.~\ref{fig:Ebind} with a vertical line indicating the initial coordinate
separation of the black holes.  We can immediately see that the dry
``quasi-circular'' R1 case has some non-zero eccentricity as the imposed
separation does not lie at the minimum of the curve. We 
also observe a shift of the minimum inwards as the wave strength
increases.  Since the coordinate separation in the parameter search 
is held fixed for the evolved initial data, the location of the system
along the binding energy curve with respect to its minimum is sufficient 
to see that the eccentricity of the orbit is likely increasing.  Unfortunately, 
the separation is too small to get a reliable measure of the eccentricity.  

\section{Evolutions of the BBH+TNW}\label{Effects of TNW} 
Our simulations of the \bbh{}+\tnw{} initial data
are summarized in Table~\ref{runtaball},
where the first row corresponds to the \bbh{} system without added radiation, the 
dry R1 run.  This will be our control case. 
We systematically evolved each \bbh{}+\tnw{} spacetime, varying $A$ and $\sigma$
of the \tnw{}.   We divide the results from evolving these
simulations into four subsections: the main result concerning the
merger time in $\S$~\ref{sec:mergertime}, the dynamical and radiated quantities from our runs
are in $\S$~\ref{sec:radiated}, the final spacetime quantities in $\S$~\ref{sec:finalspacetime}, and
a comparison of the gravitational waves by time-shifting in $\S$~\ref{sec:alignment}.  

\subsection{Merger Time}
\label{sec:mergertime}
The main result of adding gravitational radiation to our \bbh{} evolution
is to hasten the merger of the black holes.  With increasing
$E_\mathrm{ADM}$, the binary invariably merges
\emph{faster}.  The sixth column in Table~\ref{runtaball}
lists the differences in merger times between the dry R1 and the \bbh{}+\tnw{} runs 
given by $\Delta T = (T_0 - T_{0,\mathrm{dry}})/M\,.$  
The time, $T_0$, is given in units of the total, initial \ahz{} masses of the black holes and 
evaluated at the peak amplitude of each waveform extracted at a radius 
of $75M\,.$
The use of the waveform peak variation as a 
measure of the change in merger time agrees within a few percent in $\Delta T$ to 
the variation in the time it takes for the punctures to be separated by one grid spacing.

\begin{figure}[htb]
\begin{center}
\includegraphics[width=80mm]{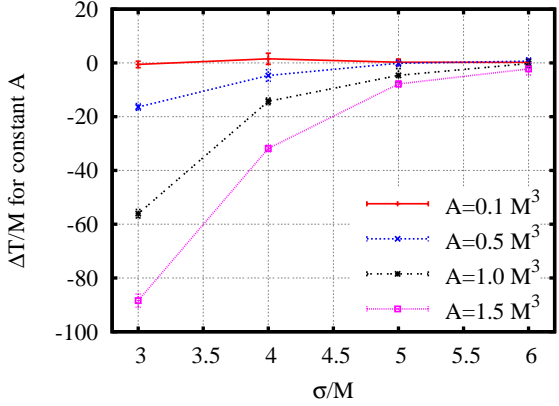}
\caption{\label{fig:constamp}Changes in merger times compared to the dry R1 run as a 
function of packet width with estimated error bars.}
\end{center}
\end{figure}

\begin{figure}[htb]
\begin{center}
\includegraphics[width=80mm]{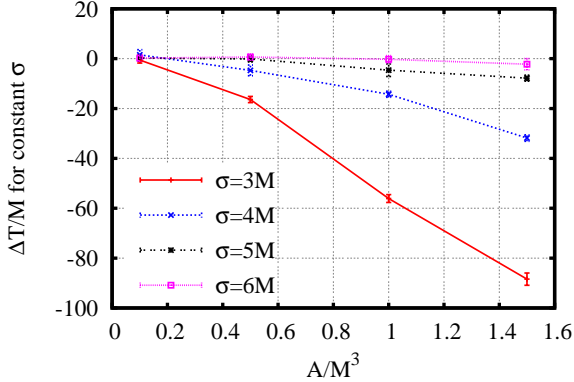}
\caption{\label{fig:constsig}Changes in merger times compared to the dry R1 run as a
function of wave amplitude with estimated error bars.}
\end{center}
\end{figure}

\begin{figure}[htb]
\begin{center}
\includegraphics[width=80mm]{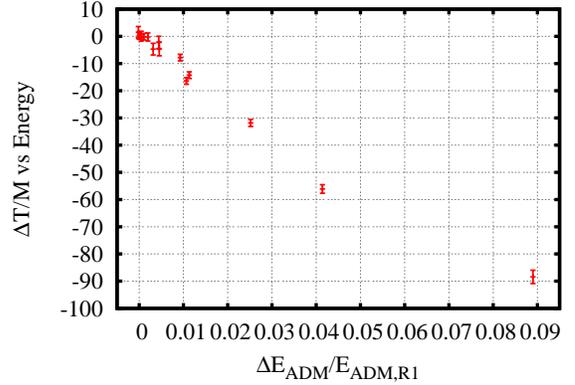}
\caption{\label{fig:dE0dtau}Changes in merger times compared to the fractional change 
in initial ADM energy of the spacetime with estimated error bars.}
\end{center}
\end{figure}

Figs.~\ref{fig:constamp} and \ref{fig:constsig} show the change in merger times from the
perspective of constant wave amplitude and constant pulse width.  We can
see that there is a strong dependence on the width of the pulse as
well as the amplitude.   Some cases show a positive value for 
$\Delta T$; however, these are all equal to zero within the errors. 
For all the $A=1 M^3$ waves that have non-zero
merger time we found an approximate power law
relation between the width of the pulse and the change in merger time:
\begin{equation} 
\Delta T(A=1 M^3) \propto \sigma^{-4.93} \,.
\end{equation} 
A more general look at the change in merger times is
found in Fig.~\ref{fig:dE0dtau}.  Given our estimated error bars,
significant changes in merger time occur when the \tnw{} has increased the initial ADM energy
of the spacetime by about 1\% compared to that of the dry R1. 

\begingroup
\squeezetable
\begin{table}
\begin{tabular}{lc||cccccc}
\multicolumn{8}{c}{Run Quantities} \\
\hline
  $A/M^3$ & $\sigma/M$ & $E_\mathrm{ADM}$ & $\frac{E_{rad}}{E_\mathrm{ADM}}$  & $\frac{J_{rad}}{J_\mathrm{ADM}}$ & $\Delta T/M$ & $M_f$ & $j_f$ \\
\hline
\hline
  0.0  & 0 & 0.9957 & 0.0359 & 0.273 & 0.0   & 0.9599 & 0.682 \\
  0.1  & 3 & 0.996  & 0.0363 & 0.273 & -0.6  & 0.9600 & 0.683 \\
  0.5  & 3 & 1.007  & 0.0451 & 0.271 & -16.4 & 0.9609 & 0.682 \\
  1.0  & 3 & 1.037  & 0.0708 & 0.263 & -56.1 & 0.9635 & 0.686 \\
  1.5  & 3 & 1.084  & 0.1058 & 0.244 & -88.4 & 0.9696 & 0.693 \\
\hline
  0.1  & 4 & 0.996  & 0.0360 & 0.274 & +1.5  & 0.9596 & 0.682 \\
  0.5  & 4 & 0.999  & 0.0385 & 0.272 & -4.7  & 0.9604 & 0.682 \\
  1.0  & 4 & 1.007  & 0.0463 & 0.272 & -14.3 & 0.9603 & 0.683 \\
  1.5  & 4 & 1.021  & 0.0589 & 0.270 & -31.8 & 0.9607 & 0.683 \\
\hline
  0.1  & 5 & 0.996  & 0.0360 & 0.273 & +0.2  & 0.9599 & 0.682 \\
  0.5  & 5 & 0.997  & 0.0369 & 0.273 & -0.1  & 0.9599 & 0.682 \\
  1.0  & 5 & 1.000  & 0.0399 & 0.272 & -4.6  & 0.9603 & 0.683 \\
  1.5  & 5 & 1.005  & 0.0448 & 0.272 & -7.8  & 0.9599 & 0.686 \\
\hline
  0.1  & 6 & 0.996  & 0.0359 & 0.273 & +0.2  & 0.9599 & 0.682 \\
  0.5  & 6 & 0.996  & 0.0364 & 0.273 & +0.7  & 0.9599 & 0.682 \\
  1.0  & 6 & 0.998  & 0.0377 & 0.273 & -0.3  & 0.9601 & 0.682 \\
  1.5  & 6 & 1.000  & 0.0399 & 0.272 & -2.2  & 0.9602 & 0.682 \\
\hline
\end{tabular}
\caption{\label{runtaball} The first two
columns are the parameters of the \tnw{s} followed by the ADM energy
of the initial spacetimes.  Column 4 and 5 give the fraction of the 
ADM energy and angular momentum radiated over the simulation.  Column 
6 is the change in merger time calculated by the shift in extracted 
waveform peak in units of the total \ahz{} mass in the initial spacetime.
Column 7 lists the final mass and Column 8 the final spin, $j_f=a_f/M_f$ of the black hole.}
\end{table}
\endgroup


To isolate how much of $\Delta T$ is due to the additional spurious radiation 
introduced and how much is due to other factors, we 
perform a series of tests.  We focus on the most
significant sources of errors, namely the resolution of our grid, wave extraction radius,
the change in proper distance in setting up the initial data, and the change in 
mass of the two black holes.  We will look at each of these factors
and assess their individual contribution to $\Delta T$.

{\em 1. Resolution:} The finest resolution for the simulations we present in Table~\ref{runtaball} is
$M/38.4$.  We check the error due to the resolution by repeating several cases
with finest resolutions of $M/44.8$ and $M/51.2$. 
We ran convergence tests on the strongest wave ($A=1.5 M^3$, $\sigma=3M$) and the 
weakest wave ($A=0.1 M^3$, $\sigma=6M$).  A third, medium, wave with $A=0.5 M^3$ and $\sigma=4M$ 
was run at just one more resolution, $M/44.8$.  For all three cases, the merger time
\emph{decreased}.  The merger time of the compact
wave decreased little for a total of $0.1M$ over the three resolutions while the 
diffuse wave decreased more drastically for a total of about $0.5M\,.$  The medium
case had a $0.2M$ difference between the two resolutions.  

\begin{figure}[tb]
\begin{center}
\includegraphics[width=80mm]{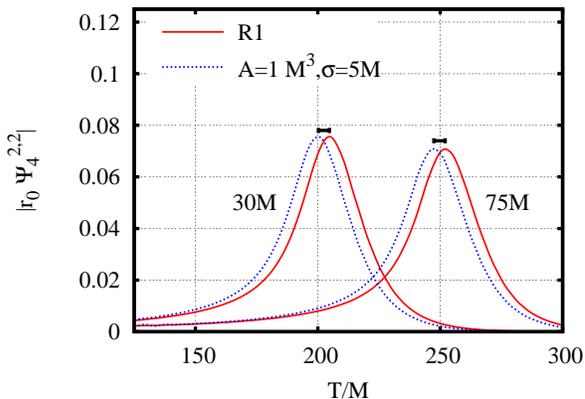}
\caption{\label{fig:gaugeeff}Sample comparison of waveforms extracted at different radii. Plotted 
are the waveforms for the dry R1 run and $A=1 M^3$, $\sigma=5M$ run extracted at $30M$ and $75M$.}
\end{center}
\end{figure}

{\em 2. Extraction Radius:} The next source of error is wave extraction radius.  In \nr{}, waveforms are usually 
calculated in terms of the Newman-Penrose scalar, $\Psi_4(t,x,y,z)$, which are extracted on a
sphere at a finite radius some distance from the source, then expanded into angular modes via 
the spin-weighted spherical harmonics, ${}_{-2}Y_{\ell m}(\theta,\phi)$.
With a proper choice of tetrad, this scalar 
is a measure of outgoing gravitational radiation.  There has been recent work 
investigating the effects the choice of extraction radius can have on the correctness of 
the waveform \cite{pazos-2006,lehner-2007}.  As the extraction radius increases, the 
errors caused by an incorrect tetrad and finite distance diminish.  While it is still an 
open question whether or not there are observable effects from the methods groups 
currently use to extract the waveforms, the methodology of the extraction is not thought 
to contaminate the waveform.  An indication the appropriate tetrad is being approached
is that the waveform amplitude scales as $1/r$, which we tested.
To get a rough estimate of the errors due to extracting at a finite radius, we compute 
$\Delta T$ using radiation extracted at $30M$ and extracted at $75M$.  In 
Fig.~\ref{fig:gaugeeff} we plot the amplitude of the dominant waveform mode, 
$|\Psi_4^{2,2}|$, extracted at the two radii for both the dry R1 run and one where the 
merger time changed significantly.  The merger time shift changed by no more than $0.4M$ 
between the two extraction radii.

{\em 3. Black Hole Mass:} Aside from unphysical sources of error, the small differences in the initial
data also change the merger time.  Though we kept the initial \ahz{}
masses nearly identical, there is still a variation of up to 0.04\% compared to
the dry R1 run.  While conducting the research for this paper, we found
that a 0.14\% change in initial \ahz{} masses of the punctures resulted in a change in merger
time of $6.7M$.  Although we do not include simulations with such a large deviation of masses,
we used this knowledge and assumed the change in merger time was linear in the change in initial mass
to estimate an error.

{\em 4. Proper Separation:} As mentioned in $\S$~\ref{Initial Data}, the presence of the additional 
gravitational radiation also increased the proper separation, $L$, from the dry R1 $9.94M$ by 
up to 2.9\%. We studied the effect of this change by evolving 
two \bbh{} spacetimes with the same initial masses and angular momentum but increasing 
the coordinate to yield proper separations of $10.24M$ and $10.10M$.  The merger
time changed by at most $2.1M$.  We assumed a linear relationship between the
$\Delta T$ and $\Delta L$ in estimating the errors from this source at each data point. 

The error bars presented in our figures are calculated by adding all the errors in quadrature:    
\begin{subequations} \label{eq:Errs}
\begin{eqnarray}
  \Sigma^2_{\Delta T}&=&\Sigma^2_T +\Sigma^2_{T_0}\, \\
  \Sigma^2_T&=& (\frac{\Delta T}{\Delta L} \Delta L)^2 + (\frac{\Delta T}{\Delta m} \Delta m )^2 + \Sigma^2_\mathrm{res} + \Sigma^2_\mathrm{tet}\,, \nonumber
\end{eqnarray}
\end{subequations}
where $\Sigma_{\Delta T}$ is the error in $\Delta T$, $\Sigma_{T}$ is the error in $T$, and $\Sigma_{T_0}$ is the error in $T$
for the dry R1 run.  $\Sigma_\mathrm{res}=0.75$ and $\Sigma_\mathrm{tet}=0.4$ are the largest measurements
for the resolution and extraction radius errors.
The accumulated errors do not account for the observed $\Delta T$ when $E_\mathrm{ADM}/E_\mathrm{R1} > 0.01\,$
and we note that the errors grow as the amplitude increases and the width decreases, most notably 
the errors associated with changes in the irreducible masses. 

The parameter space of adding spurious radiation is large.  In
Table~\ref{oddruntab}, we present the results from a few evolutions outside of our main
parameter survey.  The junk radiation present in the initial data of a typical \bbh{} simulation
may not be well represented by an $\ell=m=2$ mode.  Similarly the effect of the junk radiation might
be sensitive to the wavelength of the pulse.  In order to test how important a modulation
in the frequency might be to our conclusions,  we briefly
investigated an Eppley  packet modulated with a cosine wave, given by
\begin{equation} 
F(u)= A \cos{(k u)} u e^{-u^2 / \sigma^2} \,.
\end{equation}
This modulation adds an extra parameter controlling the wavelength of the perturbation.
We adjusted the amplitude of the wave to keep the energy approximately comparable to our standard runs.
The resulting simulation merger time differed from the unmodulated packet by less than $1M$ in $T$, well
within error bars. While this is still an avenue open to investigation, we concluded
that the modulation was not affecting the results enough to warrant an additional parameter in our survey.
We also conducted a test of the geometry of the wave by 
initiating a pulse with an $\ell=2$, $m=0$ mode.  Again we changed the amplitude so that
the energy in the wave was approximately constant and found that there was a change
in merger time of $1M$ compared to the $\ell=2$, $m=2$ simulation, again within error bars.  
This points towards the wavelength and angular dependence
of the pulse being secondary to the additional energy in determining the effect of 
the pulse on the merger time.  

Finally, to make a stronger connection to the junk 
radiation being associated with each puncture, we 
added \emph{two} identical waves centered at each of the black holes rather
than at the center.  Compared to the same wave initiated at the center, the dual waves added
almost twice the energy and almost doubled the change in merger time, which is consistent with the
center-of-mass \tnw{}. 

\begingroup
\squeezetable
\begin{table}

\begin{tabular}{cccc|cccc}
  $A/M^3$ & $\sigma/M$ & $Mk$ & $m$ & $\frac{E_{rad}}{E_{\mathrm{ADM},i}}$ & $\frac{J_{rad}}{J_{\mathrm{ADM},i}}$ & $M_f$ & $\Delta T/M$ \\
\hline
\hline
  1 & 3 & 0 & 2 & 0.0696 & 0.252 & 0.970 & -65.5 \\
  0.25 & 3 & 0 & 2 & 0.0382 & 0.272 & 0.970 & -5.6  \\
  0.15 & 3 & 0 & 0 & 0.0382 & 0.271 & 0.961 & -6.9  \\
  $3\times 10^{-3}$ & 4 & 2 & 2 & 0.0363 & 0.272 & 0.962 & -4.5 \\
  $7\times 10^{-4}$ & 4 & 3 & 2 & 0.0358 & 0.272 & 0.962 & -4.0  \\
  Dual 0.5 & 4 & 0 & 2 & 0.0411 & 0.271 & 0.960 & -8.9  \\
\hline 
\end{tabular}

\caption{\label{oddruntab} Overview of the odd runs.  The left four columns are the wave parameters,
followed by the fraction of ADM energy radiated and the fraction of the ADM angular momentum radiated.  The final
masses are given in the 7th column followed by the merger time change as derived by the peak of the waveform 
extracted at $75M$.}
\end{table}
\endgroup

\subsection{Dynamical and Radiated Quantities}
\label{sec:radiated}
We now investigate the effects of the \tnw{s} on the radiated quantities
derived from the waveforms.  We calculate these quantities from the
Weyl scalar $\Psi_4$ assuming, as in the calculation of the waveform,
the fiducial tetrad of Baker et al.~\cite{Baker02}.  A summary of
the quantities obtained from $\Psi_4$ are listed in Table~\ref{runtaball}.
The fraction of the initial ADM energy radiated was calculated across a detector at $40M$.  As
expected, the radiated energy increases with the strength of the wave.
When $A<0.5M^3$ and $\sigma\ge5M,$ there is no measurable difference
between the \bbh{}+\tnw{} and the \bbh{} cases within numerical error. 
For those cases, we can only conclude that the energy in the
wave propagates out without a measurable interaction with the
black holes. Similarly a trend emerges as we increase $A$ for each $\sigma$, which corresponds
to increasing $E_\mathrm{ADM}$.  The radiated angular momentum consistently decreases.

\begin{figure}[hb]
\begin{center}
\includegraphics[width=80mm]{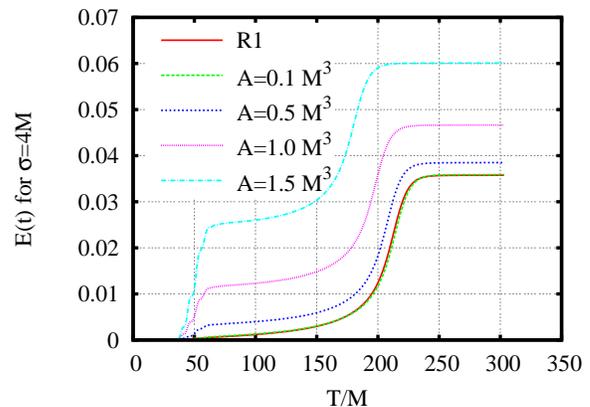}
\caption{\label{fig:et50}Energy radiated across a sphere of radius $r=40M$ as calculated from the Weyl
scalar $\Psi_4$.}
\end{center}
\end{figure}

To look at the interaction of the \tnw{} with the black holes as it propagates out, we 
study the radiated energy and angular momentum as functions of time.
In Fig.~\ref{fig:et50} we plot the energy radiated across a detector at radius $r=40M$.  We see the
energy grows from a time of $40M$ to around $80M$ as the initial burst
of spurious radiation passes the detector.  After this burst of energy
the remaining energy radiated is approximately $0.035E_{\mathrm{ADM},\mathrm{R1}}$ and is almost
uniform across the various cases.  
From this we can see that most of the energy introduced in the spacetime is
quickly flushed out of the system, leaving a system which radiates a
further amount of energy that is independent of the junk radiation.

\begin{figure}[ht]
\begin{center}
\includegraphics[width=80mm]{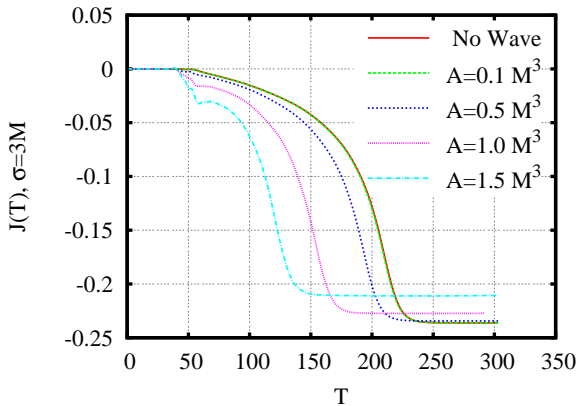}
\caption{\label{fig:jt50}Angular momentum radiated across a sphere of radius $r=40M$ 
for $\sigma$ of $3M$ as calculated from the Weyl scalar $\Psi_4$.}
\end{center}
\end{figure}

\begin{figure}[h]
\begin{center}
\includegraphics[width=80mm]{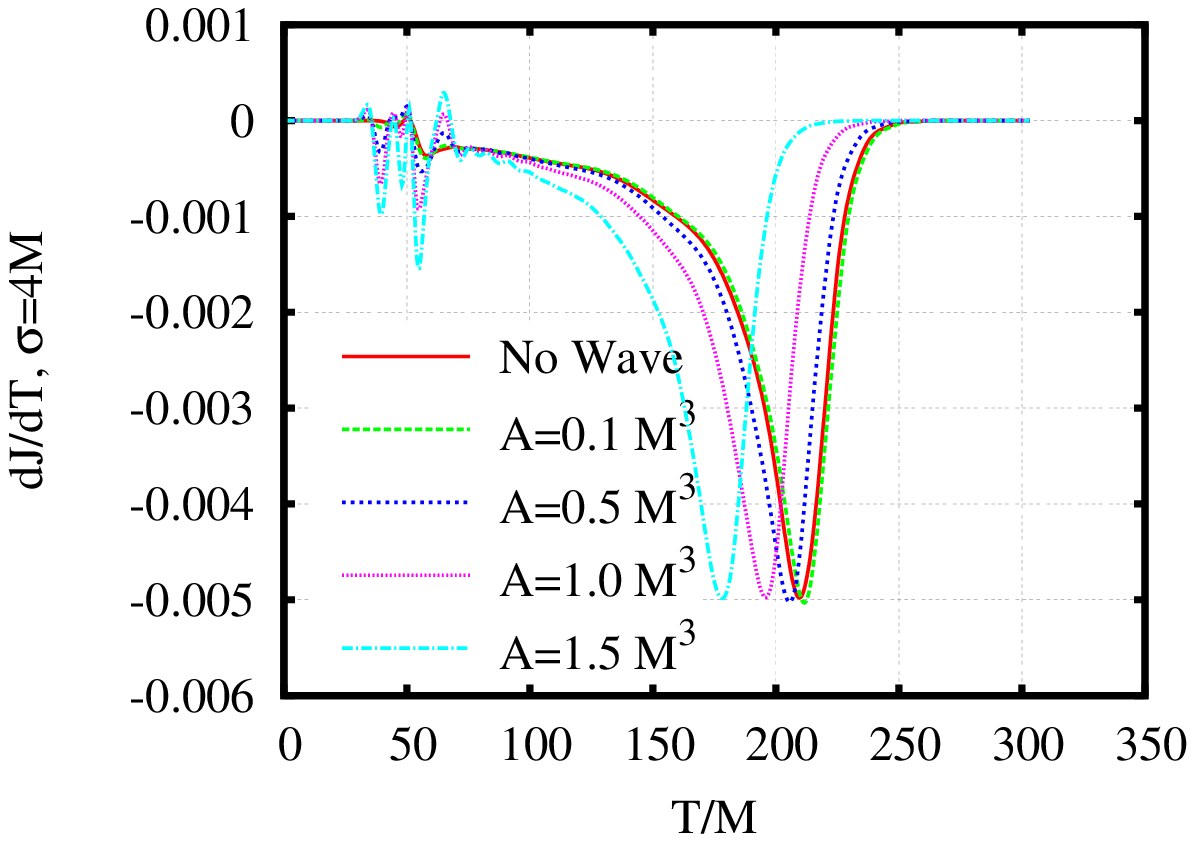}
\caption{\label{fig:djdt}Flux of angular momentum across a sphere of radius $r=40M$ 
for $\sigma$ of $4M$ as calculated from the Weyl scalar $\Psi_4$.}
\includegraphics[width=80mm]{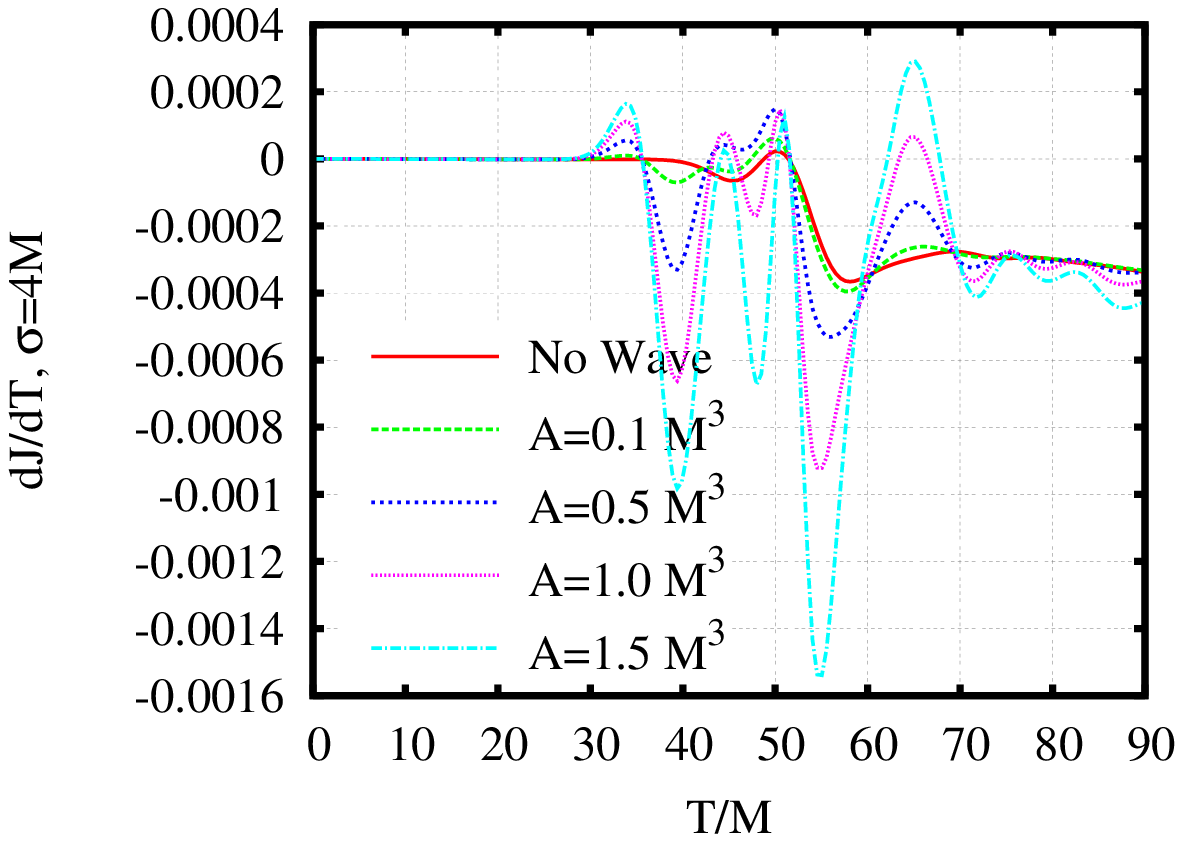}
\caption{\label{fig:djdtclose}Closeup of angular momentum flux across a 
sphere of radius $r=40M$ concentrating on what the spurious radiation carries.}
\end{center}
\end{figure}

All the evolutions started from spacetimes with equal $J_\mathrm{ADM}$ since the
\tnw{} does not add angular momentum to the \bbh{} spacetime.
In column five of Table~\ref{runtaball} and in Fig.~\ref{fig:jt50}, we see that the amount 
of angular momentum radiated across a detector located at $40M$ is independent of the wave
with some numerical error.
The difference in $J_{rad}$ between the runs lies in \emph{when} the system radiates
the angular momentum.  This is better seen in Fig.~\ref{fig:djdt} where we present the angular momentum 
flux across the sphere at $r=40M$ and in Fig.~\ref{fig:djdtclose}, 
a close-up of the initial part of the data. The spurious radiation is transporting extra angular
momentum as it is flushed out. 


Figs.~\ref{fig:et50} and \ref{fig:jt50} show some of the \tnw{} propagating
out at the same time as the spurious radiation is flushing out.
One of the consequences of the wave traveling away from the center of 
the orbit is that it can interact with the black holes and potentially 
increase the mass of each black hole during the early part of the inspiral. 
Table~\ref{mirrtab} documents how the \ahz{} masses change
as a function of $A$ and $\sigma$ for the stronger \tnw{} cases.  The black-hole mass
is calculated using an \ahz{} tracker \cite{thornburg}. The $\Delta m$ is a measure
of the change in the mass of each black hole up to $50M$, such that $\Delta m = m(t=50M) - m(t=0)\,.$
The change in the initial ADM energy compared to the R1 run is given by
$\Delta M = E_{\mathrm{ADM}} - E_{\mathrm{ADM},\mathrm{R1}}\,$ at $t=0\,.$
We use this estimate of the differences in the mass of the 
spacetime between R1 and the rest of the runs to compute a naive estimate of 
the total fraction of energy absorbed by both black holes.  Up to 7.89\% of the extra ADM energy
is observed to be absorbed by the black holes during the first $50M$
of the simulation, ~3.9\% by each black hole. The actual amount absorbed 
depends strongly on the width of
the wave: the narrow, strong pulses are more readily absorbed than
the weak, diffuse pulses that extend beyond the black holes
in the initial data.  In the weaker cases the change is barely visible
above the noise in the \ahz{} mass calculation, in
the stronger cases it is unmistakable. 

To assess how important the absorption of energy by the black holes during the evolution is to the 
changes in merger time and radiated angular momentum, we refer to our discussion about the sensitivity 
of the merger time to a change
in the initial \ahz{} masses in $\S$~\ref{sec:mergertime}.  Given a change in mass of the individual black holes
of 0.14\%, the merger time changed by $6.7M$.  In setting up the initial data,
we do not allow the \ahz{} masses to change more than 0.04\%.  The amount of absorption measured 
during the evolution is as much as 3.9\%; and, therefore, the increase in mass 
may be accounting for some, although not all, of the effects of the \tnw{}.
The outliers, the cases of most
extreme merger times, merge so quickly that differentiating the burst
of spurious radiation and region of pure inspiral is difficult.  
We also tried normalizing the time axis by the total \ahz{} masses after the wave has passed rather than
that from the initial spacetime.  This changed $\Delta T$ by no more than $1M$ so the choice of normalization
does not account for the observed difference in merger times.

\begingroup
\squeezetable
\begin{table}
\begin{tabular}{cccccc}
\multicolumn{6}{c}{Masses} \\
\hline
  $A/M^3$ & $\sigma/M$ & $\Delta m/M$ & $\Delta M/M$ & $2\Delta m /(\Delta M)$ & $M_f/(2m(t=50))$ \\
\hline
\hline
  0  & - & 0 & 0 & 0 & 0.950 \\
\hline
  0.5  & 3 & $3.90\times {10}^{-4}$ & 0.0106 & $7.34\times {10}^{-2}$ & 0.950 \\
  1.0  & 3 & $1.56\times {10}^{-3}$ & 0.0412 & $7.57\times {10}^{-2}$ & 0.951 \\
  1.5  & 3 & $3.49\times {10}^{-3}$ & 0.0886 & $7.89\times {10}^{-2}$ & 0.953 \\
\hline
  1.0  & 4 & $2.84\times {10}^{-4}$ & 0.0113 & $5.03\times {10}^{-2}$ & 0.950 \\
  1.5  & 4 & $6.41\times {10}^{-4}$ & 0.0251 & $5.11\times {10}^{-2}$ & 0.949 \\
\hline
  1.0  & 5 & $5.89\times {10}^{-5}$ & 0.0045 & $2.62\times {10}^{-2}$ & 0.950 \\
  1.5  & 5 & $1.26\times {10}^{-4}$ & 0.0092 & $2.74\times {10}^{-2}$ & 0.950 \\
\hline
  1.0  & 6 & $1.76\times {10}^{-5}$ & 0.0020 & $1.76\times {10}^{-2}$ & 0.950 \\
\hline
\end{tabular}
\caption{\label{mirrtab} Change in \ahz{} mass compared to
the difference in initial ADM energy for the stronger waves.  $\Delta m$
is the change in a single black hole AH mass over the first $50M$,
$\Delta M$ is the additional ADM energy compared to the 
dry R1 run. Column 5 is the fraction of the
extra ADM energy absorbed by both black holes combined, and the
last column is the ratio of the final black hole mass to the total
\ahz{} mass after the wave has been absorbed.}
\end{table}
\endgroup


\subsection{Final Spacetime}
\label{sec:finalspacetime}
One of the important products of a \bbh{} coalescence to relativists
and astrophysicists are the final black hole's mass and spin.
The final black-hole masses and spins are presented in 
the last two columns of Table~\ref{runtaball}.  To compute
the final mass, $M_f$, we use energy conservation arguments by calculating
the difference between the ADM energy and the radiated energy as calculated from the Weyl scalar
$\Psi_4\,.$  The final spin, $j_f=a_f/M_f$, is calculated by finding the
complex ringdown frequency in the $\ell=2$, $m=2$ mode and using the
numerical Kerr frequencies given in Table~II of Appendix D in Berti et
al~\cite{Berti05} to find the corresponding spin parameter.  This method
agrees within stated errors to inverting the fit of Eq.~(E2) of the same paper.  Given the
strong dependence of the spin on the damping time, we limit ourselves
to the real part of the complex frequency and compare this to a
separate spin calculation using the isolated horizon framework~\cite{Dreyer} where possible.  

From the values of $M_f$ and $j_f$ listed in Table~\ref{runtaball},
we can see that the final spins are constant within numerical accuracy and the final masses do not 
vary strongly with $A$ and $\sigma$. 
The trend is an increase in the final mass with increasing $E_\mathrm{ADM}$
becoming noticeably greater than our numerical errors for the four largest cases,
$M_f \ge 0.963 \,.$  From this we can see that the narrower pulses not only have
more energy, but they also interact more efficiently with the black holes.
Being more readily absorbed by the punctures, they increase the individual masses and thus the final
mass. 
The last column of Table~\ref{mirrtab} shows the ratio of the final mass to the 
total \ahz{} mass once the wave has interacted with the inspiraling black holes.  We see 
the ratio is roughly constant, implying that approximately 5\% of the initial \ahz{} mass 
is radiated away \emph{if} we include the wave energy absorbed by the black holes.  The 
exception is the most extreme wave where the black holes merged before all the spurious radiation has 
been absorbed into the \ahz{}.  This would underestimate the \ahz{} growth and thus overestimate 
the value of the ratio.
The change in final mass agrees within numerical error to the change in 
total \ahz{} mass after the wave has interacted with the inspiraling black holes except for the case 
of the strongest wave.  In that case the \ahz{s} have not absorbed all the energy before 
the black holes begin to merge so we are underestimating the growth of these
black holes.  

\subsection{Alignment of Amplitude and Phase}
\label{sec:alignment}
\begin{figure}[htb]
\begin{center}
\includegraphics[width=80mm]{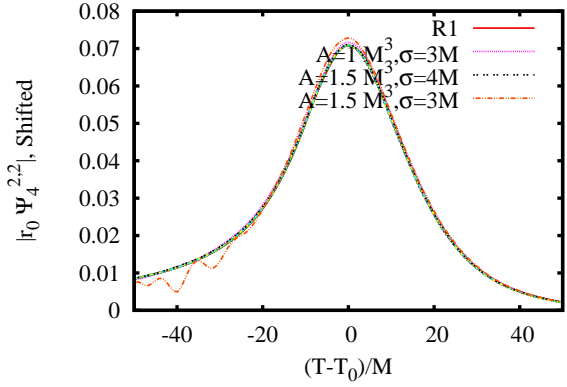}
\caption{\label{fig:shiftamp}Comparison of $|\Psi_4^{2,2}|$ for all runs shifted such that the peak amplitudes
before ringdown coincide. This lets us compare the relative damping times of the ringdown and thus
the properties of the final black hole.  We also note the agreement for about $50M$ before ringdown as well. 
Though the legend only labels the distinguishable cases, all the runs are contained in the figure.}
\end{center}
\end{figure}

\begin{figure}[htb]
\begin{center}
\includegraphics[width=80mm]{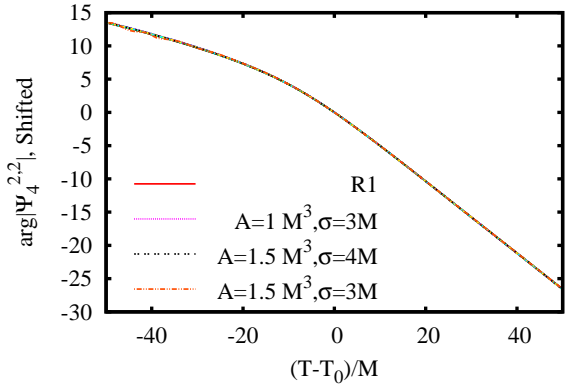}
\caption{\label{fig:shiftph}Comparison of $\mathrm{arg}(\Psi_4^{2,2})$ for all runs shifted such that the peak amplitudes
before ringdown coincide and shifted vertically.  This lets us compare the relative frequencies of
ringdown. Though the legend only labels the distinguishable cases, all the runs are
contained in the figure.}
\end{center}
\end{figure}

As stated in the introduction, a common method to compare
waveforms is via a time-shift of the
amplitude of each waveforms such that their peaks overlap, the result
of which is shown in Fig.~\ref{fig:shiftamp}.  We can see that the
waveforms overlap very well after the merger.  The only noticeable
difference is for the strongest \tnw{} we evolved on the \bbh{} system,  
the $A=1.5M^3$, $\sigma=3M$ case, where we find the largest difference in the final black hole
compared to the dry R1 run.  
We can also see residual contamination of the merger portion of the waveform by the
spurious radiation due to the binary merging so quickly.  Similarly, we 
shift the waveform phase such
that they overlap at $T=T_{\mathrm{peak}}$ in Fig.~\ref{fig:shiftph}.
The agreement in the phase's slope during ringdown is further confirmation that the mass and spin
of the final black hole are not significantly altered.

The waveform overlap in the merger
regime continues to start before the merger, as seen in \cite{Baker07}, as long as 
the spurious radiation does not contaminate this region of the waveform.  This alludes to the relatively
simple form of the merger waveform seen in all the various situations
currently tested.  As long as the spurious radiation is not strong
enough to noticeably alter the final black hole, the merger portion of
the waveform remains essentially unaltered and the contamination
to the system predominately results in the change in merger time and thus
a time-shift of the waveform.


\section{Conclusions}\label{Conclusions}
In this study, we simulated an equal-mass, non-spinning \bbh{} system through 
its last orbits, merger and ringdown.  The system is perturbed by the systematic
addition of spurious radiation in the form of a Teukolsky-Nakamura gravitational wave at the 
binary's center of mass.  The initial energy of the wave is tunable, specified
by the amplitude and width of the radiation; in addition, the initial angular momentum 
was fixed for the entire sequence of runs. The binaries that contain the extra radiation 
invariably merge faster than those with no additional radiation. 

In addition to the main result of decreasing merger time, 
some changes to the radiated quantities and the final black hole were measurably above numerical
error.  This occurred once the additional energy provided by the \tnw{} was equal to or greater than 1\% of the 
dry \bbh{} spacetime.   As the \tnw{} propagated out of the center,  approximately 4\% 
of additional ADM energy was absorbed by each black hole.  In that strong-wave case,
it was not possible to make an accurate measurement of the mass of the enlarged black holes before the
plunge of the binary.  The final spins of the black holes, however, remained unaffected
by the gravitational radiation for all but the strongest case ($A=1.5M^3$, $\sigma=3M$).  
The constant black-hole spin
is consistent with the wave slightly increasing the eccentricity of the orbit 
for small eccentricities \cite{Hinder:2007qu,Sperhake-2007}.  We also observed a decrease in the radiated angular momentum
with increasing \tnw{} strength.

We conjecture, based on the change in the initial binding energy of the \bbh{}+\tnw{} systems and Newtonian 
back-of-the-envelope calculation, that the spurious radiation increases the eccentricity of the original orbit.
Unfortunately, the separation of the black holes was not large enough to enable a reliable calculation 
of the eccentricity.  The merger time is very sensitive to the increase in individual black-hole
masses via wave absorption; however, this was not enough to account for the observed change in the 
time of merger even when ignoring the strongest wave case.  
The combined effects of increasing the individual black-hole masses and the eccentricity
of the orbit caused the binaries to merge faster with increasing energy.

One of the conjectures in the literature is that the spurious radiation, 
intrinsic to the construction of initial data for \bbh{} evolutions, is flushed out of the
simulation within a crossing-time and does not effect the radiation or the binary.
We can relate the results of this study to other \bbh{} evolutions by looking at 
the early changes in \ahz{} mass as well as how much 
energy leaves the system in the burst of spurious radiation.  For the dry R1 run, the energy radiated in 
the initial pulse is $9 \times 10^{-4} E_\mathrm{ADM,R1}$.  We find that there
is negligible effect on the merger time at that level.  
Our results indicate that the spurious
radiation present in initial data sets is unlikely to cause dramatic departures from the 
true \bbh{} solution and therefore we can state that the simulated merger is robust to the 
presence of spurious radiation.

\acknowledgments
This work was supported in part by NSF grants
PHY-0653443, PHY-0555436 and PHY-0114375
(CGWP).  Computations were performed at NCSA and TACC under allocation
TG-PHY060013N, and at the Information Technology Services at Penn State.
The authors thank M.~Ansorg, E.~Bentivegna, 
A.~Knapp, P.~Laguna, R.~Matzner, E.~Schnetter, U.~Sperhake and J.~Thornburg for contributions to the
computational infrastructure and helpful discussions, and E.~Berti for
the data tables used in the quasi normal fitting.

\bibliography{references}  

\end{document}